\documentclass[parallelisme]{compas2014}
\usepackage{amssymb,amsmath}
\usepackage[T1]{fontenc}
\usepackage[utf8]{inputenc}
\usepackage{amsmath,amsfonts,amssymb}
\usepackage{graphicx}
\usepackage{amsfonts,amsmath,amssymb,graphicx}
\usepackage{color}
\usepackage{amsthm}
\usepackage{algorithm,algorithmic}
\usepackage{subfig}
\usepackage{url}
\begin{document}
%
%
\title{Partitionnement Déterministe pour Résoudre les Problèmes de Programmation Par Contraintes en utilisant le Framework Parallèle Bobpp}
\author{
Tarek Menouer and Bertrand Le Cun\\
University of Versailles Saint-Quentin-en-Yvelines, France \\
Tarek.menouer@prism.uvsq.fr, Bertrand.lecun@prism.uvsq.fr
}

\maketitle

\begin{abstract}
Cet article présente une stratégie de partitionnement déterministe pour paralléliser le parcours des espaces de recherche, dans le but de résoudre les problèmes de Programmation Par Contraintes (PPC). Ce travail est réalisé dans le cadre d'un projet industriel nommé \textit{PAJERO}, dont le but est de proposer un solveur de PPC parallèle qui doit répondre toujours la même solution en utilisant le mode séquentiel ou parallèle.
Il est clair que l'exploration parallèle des espaces de recherche modifie l'ordre dans lequel les nœuds solutions sont visités.
Dans le contexte où la première solution trouvée est retournée à l'utilisateur, l'utilisation de plusieurs cœurs de calcul peut modifier la solution retournée.
Dans la littérature, plusieurs stratégies non déterministes ont été proposées pour paralléliser l'exploration des espaces de recherche. La plupart de ces stratégies sont basées sur la technique de vol de travail (Work Stealing). Celle-ci est utilisée pour partitionner dynamiquement l'espace de recherche en sous-espaces, puis à affecter chaque sous-espace à un cœur de calcul.
Notre étude consiste à rendre l'algorithme de recherche parallèle déterministe par rapport à l'algorithme séquentiel.
Nous considérons que l'algorithme de recherche séquentielle est déterministe. La solution proposée est simple et élégante. Elle consiste à introduire un ordre total sur les nœuds, dans le but de retourner toujours la même solution que la première solution retournée en séquentiel, quel que soit le nombre de cœurs utilisés.
Pour évaluer cette stratégie de partitionnement déterministe, nous avons effectué des expériences en utilisant un solveur de PPC nommé Google OR-Tools au-dessus de notre framework parallèle Bobpp. Les performances sont illustrées par la résolution des problèmes de PPC modélisés en utilisant le format FlatZinc.

\MotsCles{Parallélisme, Programmation Par Contraintes, Déterminisme, Vol de travail}
\end{abstract}
\section{Introduction}\label{intro}
Nous présentons dans cet article une version deterministe d'algorithme de recherche parallèle utilisé pour résoudre des problèmes d'optimisation ou de satisfaction de contraintes.
Dans la littérature, il existe plusieurs études sur la parallélisation des espaces de recherche, comme celles effectuées dans le domaine de la Programmation Par Contraintes (PPC)~\cite{IPCJIMNCPNPPCU,LP,Rolf11,LaSH09}. Il existe aussi quelques solveurs de PPC parallèles tels que Gecode~\cite{MN}, Parallel COMET~\cite{MLSAVP2007} et Parallel ILOG~\cite{ibmilog}, etc. La majorité des solveurs de PPC parallèles utilise la technique de vol de travail (Work Stealing)~\cite{GCCSPS2009} pour paralléliser l'exploration des espaces de recherche.

Dans le cadre de la Programmation Mathématique (PM), il y a aussi beaucoup de travaux effectués pour paralléliser l'exploration des espaces de recherche. Pour autant que nous sachions, il y a peu de solutions où la recherche parallèle est déterministe. Par exemple, dans le logiciel commercial CPLEX développé par IBM, l'utilisateur a le choix entre l'utilisation du mode déterministe ou opportuniste. IBM a introduit cette fonctionnalité pour répondre aux besoins de ses clients. Toutefois, il n'existe pas de publications montrant comment le déterminisme est abordé dans ce cas. 

Dans le cadre d'un projet industriel nommé \textit{PAJERO} financé par OSEO-ISI~\cite{Pajero}, un algorithme de recherche parallèle a été proposé pour explorer les espaces de recherche. Cette stratégie, appelée la Stratégie de Partitionnement Dynamique Anticipée (SPDA)~\cite{TMBLPCO13}, est basée sur une méthode de partitionnement dynamique en utilisant la technique de vol de travail.

La SPDA fonctionne bien, avec un bon équilibrage de charge entre les cœurs de calcul, pour résoudre les Problèmes de Satisfaction de Contraintes (PSC) et les Problèmes d'Optimisation de Contraintes (POC) en utilisant les techniques de la PPC. Elle est implémentée en utilisant le solveur OR-Tools~\cite{LP} au-dessus de notre framework parallèle Bobpp~\cite{GaLe07}.
OR-Tools est un solveur de PPC séquentiel, il est développé en open source par l'équipe de recherche Google.
Bobpp est un framework open source qui fournit une interface entre les solveurs de problèmes combinatoires et les machines parallèles.

Le problème de la SPDA est que l'algorithme de recherche n'est pas déterministe.
Comme pour CPLEX, où la fonctionnalité de déterminisme a été introduite pour répondre aux besoins des clients, nous avons proposé une version modifiée de la SPDA pour assurer que la solution retournée à l'utilisateur est toujours la même quel que soit le mode d'exécution séquentiel ou parallèle.

Avoir la même solution pour chaque exécution, facilite pour le programmeur le débogage pendant la phase de développement. Cela facilite également le test des programmes parallèles, en utilisant la même technique que pour les programmes séquentiels.

L'originalité de notre étude est de proposer une Stratégie de Partitionnement Déterministe (SPD) basée sur les propriétés de la recherche pour résoudre les problèmes de PPC.

D'abord, nous considérons que l'algorithme de recherche séquentielle est déterministe, ce qui signifie que le solveur séquentiel donne toujours la même solution. En utilisant le mode parallèle, la première solution ou la première solution optimale retournée n'est pas nécessairement la même que la solution retournée en mode séquentiel. En effet, pour la recherche parallèle, l'ordre dans lequel les nœuds solutions sont visités n'est pas toujours le même ordre que l'ordre obtenu en mode séquentiel.

Pour effectuer une recherche parallèle déterministe, la SPD utilise le même principe que la SPDA qui consiste à partitionner l'espace de recherche en sous-espaces pendant l'exécution de l'algorithme de recherche. Ensuite, elle affecte chaque sous-espace à un cœur de calcul.

Pour trouver toujours la même solution, nous introduisons un identifiant unique pour chaque nœud solution. L'identifiant de nœud représentant le chemin du nœud racine au nœud solution.
A chaque fois qu'un cœur de calcul trouve une solution, il sauvegarde cette solution en tant que la solution la plus à gauche et arrête tous les autres cœurs qui travaillent à droite de la branche de la solution la plus à gauche. La solution la plus à gauche est partagée entre tous les cœurs de calcul. Quand une nouvelle solution est trouvée, nous comparons l'identifiant de cette nouvelle solution avec l'identifiant de la solution la plus à gauche. L'identifiant de cette nouvelle solution est forcément plus à gauche que l'identifiant de la solution la plus à gauche, dans ce cas, nous mettons à jour la solution la plus à gauche avec cette nouvelle solution. Selon ce principe, la solution la plus à gauche est la même que la première solution trouvée en séquentiel.

Ce principe est appliqué seulement pour résoudre les PSC, car toutes les solutions ont la même évaluation. Par contre, pour résoudre les POC, les nœuds solutions n'ont pas tous la même évaluation. Par conséquent, l'objectif dans la résolution des POC est de trouver la solution la plus optimale. Afin de retourner toujours la même solution optimale, nous proposons qu'à chaque fois qu'un cœur de calcul trouve une nouvelle solution, il ne doit pas arrêter les autres cœurs, et il met à jour la solution optimale la plus à gauche si :

\begin{itemize}
\item L'évaluation de la nouvelle solution est plus optimale que l'évaluation de la solution optimale la plus gauche,
\item L'identifiant de la nouvelle solution est plus à gauche que l'identifiant de la solution optimale la plus à gauche et les deux solutions ont la même évaluation.
\end{itemize}

La section suivante présente la définition du déterminisme dans le parallélisme. Dans la section~\ref{SPD}, le principe de la SPD est présenté. Des expérimentations pour résoudre des problèmes de PPC modélisés en utilisant le format Flatzinc sont données dans la section~\ref{xp}. Enfin, une conclusion et quelques perspectives sont proposées en section~\ref{conclusion}.
\section{Définition du Déterminisme dans le Parallélisme}\label{de_pa}

Un problème de PPC est constitué d'un ensemble de variables, $X = \{x_1, x_2, ..., x_n\}$, pour chaque variable $x \in X$, il existe un ensemble fini de domaines de valeurs, $D(x) = \{a_1, a_2, ..., a_k\}$ et une collection finie de contraintes, $C = \{c_1, c_2, ..., c_m\}$.

Chaque contrainte $c_i \in C$ contient un certain ensemble de variables Vars(C), et elle peut être considérée comme étant un sous ensemble du produit des domaines de variables de Vars(C) (i.e $C$ est l'ensemble des \textit{Tuples} qui satisfont les contraintes).

Un problème ($\beta$) de PPC peut être résolu comme suit :
Au début, toutes les variables de $\beta$ sont non assignées. A chaque étape, une variable $x$ est choisie, et une valeur possible $a \in D(x)$ est affectée à son tour. Chaque branche d'un arbre de recherche calculée par cette recherche définit une affectation. Ensuite, le mécanisme de propagation vérifie, pour chaque valeur la cohérence de cette affectation partielle avec les contraintes. En cas de cohérence, un appel récursif est effectué. Chaque affectation partielle crée un nœud dans l'arbre de recherche. Ainsi, nous associons la consistance d'un nœud avec la cohérence implicite d'une affectation.

Il existe plusieurs heuristiques pour choisir les variables de branchement, et pour chaque variable la valeur à affecter, ces heuristiques sont appelées les méthodes de branchement~\cite {WM04} et~\cite{11HYSM09}.

Les méthodes de branchements peuvent dépendre de :

\begin{itemize}
\item Les données locales d'un nœud : Le branchement ne dépend pas de l'historique de la recherche et de l'exploration de l'espace de recherche. Dans ce cas, nous pouvons dire que l'algorithme de recherche séquentielle qui utilise cette méthode de branchement est un algorithme déterministe et la première solution trouvée en séquentiel est représentée par la solution la plus à gauche pour chaque exécution.

\item L'apprentissage des informations : Il est possible de relancer la recherche à chaque fois qu'une nouvelle information sur l'espace de recherche est obtenue. Dans ce cas, ni l'algorithme séquentiel ni l'algorithme parallèle ne peut être un algorithme déterministe.

\item Le branchement aléatoire : Le choix des variables et le choix des valeurs de chaque variable sont totalement aléatoires. Dans ce cas aussi, nous ne pouvons pas avoir un algorithme déterministe.
\end{itemize}

Heureusement, le solveur OR-Tools utilise une heuristique déterministe pour choisir le branchement des variables sur la base des données locales des nœuds. OR-Tools est un solveur de PPC développé en C++ par l'équipe de recherche Google. Le principe de ce solveur est d'explorer l'arbre de recherche afin de trouver une ou toutes les solutions possibles en utilisant la technique de Backtrack~\cite{BJRE75}.

En utilisant le déterminisme dans le parallélisme, la performance peut être modélisée en prenant une entrée (problème) et en calculant une sortie (solution). Dans cette étude, la sortie est déterminée uniquement par les entrées des données et non pas par le temps de calcul, l'accélération, l'ordonnancement des cœurs ou le nombre de cœurs utilisés.

Le programme parallèle est déterministe si, pour une entrée donnée, chaque exécution du programme parallèle produit un résultat identique et ce résultat est le même que le résultat obtenu en utilisant le mode séquentiel. Dans la littérature, il existe quelques travaux réalisés pour rendre les algorithmes déterministes~\cite{BAAS09}, \cite{OMAJAS09} et \cite{ DJLBCLOM09}.

\begin{figure}
\center \includegraphics[scale=0.45]{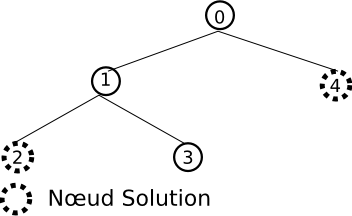}
\caption{Arbre de recherche}
\label{fig_arbre1}
\end{figure}

La figure~\ref{fig_arbre1} représente un arbre de recherche avec deux solutions. Pour trouver la première solution, on peut utiliser deux modes d'exécution :

\begin{itemize}
\item Le mode séquentiel : le cœur 0 visite le nœud 1 ensuite le nœud 2 qui est un nœud solution,
\item Le mode parallèle en utilisant deux cœurs de calcul : le cœur 0 visite le nœud 1 et le cœur 1 visite le nœud 4 qui est un nœud solution.
\end{itemize}

En utilisant le mode séquentiel, la première solution trouvée est représentée par le nœud 2. En utilisant le mode parallèle, la première solution trouvée est représentée par le nœud 4. En conséquence, l'ordonnancement parallèle et séquentiel ne donnent pas toujours la même solution. Dans la section suivante, nous présentons une nouvelle stratégie de partitionnement déterministe parallèle qui donne toujours la même solution quel que soit le mode d'exécution (parallèle ou séquentiel) et quel que soit le nombre de cœurs de calcul utilisés par le mode parallèle.
Comme actuellement la majorité des logiciels sont composés de modules complémentaires entre eux, l'utilisation d'un algorithme déterministe peut faciliter aux programmeur la parallélisation d'un module sans modifier le fonctionnement global du logiciel. Comme présenté dans ~\cite{BAAS09}, l'utilisation d'un algorithme déterministe présente d'autres avantages aux programmeurs, par exemple :

\begin{itemize}
\item Faciliter le débogage des programmes au cours de développement de son logiciel,
\item Tester le programme parallèle en utilisant la même technique que pour le programme séquentiel.
\end{itemize}

\section{Stratégie de Partitionnement Déterministe (SPD)}\label{SPD}

Le but de cette stratégie est de proposer un algorithme parallèle déterministe. 
Pour effectuer une recherche parallèle, la SPD utilise le même principe que la SPDA en utilisant le solveur OR-Tools au-dessus du framework parallèle Bobpp.

Le framework Bobpp est utilisé comme un support d'exécution. L'objectif de ce framework est de fournir un environnement unique pour la plupart des classes de problèmes d'optimisation combinatoire qui peuvent être résolus en utilisant différents environnements de programmation comme les POSIX threads ainsi que MPI~\cite{GaLe07}.

Le principe utilisé par la SPDA est basé sur une méthode de décomposition dynamique de l'arbre de 
recherche. 
A chaque fois qu'un cœur de calcul visite un nœud dans l'arbre de recherche, il teste s'il n y a pas un cœur en attente. Dans ce cas, le cœur actif, qui explore les nœuds, arrête la recherche sur le nœud OR-Tools le plus à gauche. Le chemin du nœud racine jusqu'au nœud droit est enregistré dans ce que nous appelons le \textit{nœud-BOB}. Le \textit{nœud-BOB} est inséré dans la File de Priorité Globale de Bobpp appelée \textit{FPG}, ensuite le cœur actif, poursuit la recherche avec le nœud OR-Tools gauche. Sinon, s'il n'y a pas de cœur en attente, le cœur actif effectue une recherche locale à l'aide du solveur OR-Tools. Cette stratégie de partitionnement est expliquée en détail dans~\cite{TMBLPCO13}.

Pour des raisons techniques dans le solveur OR-Tools, à chaque fois qu'un cœur de calcul commence une nouvelle recherche en utilisant un \textit{nœud-BOB}, il est obligé d'effectuer une recherche redondante du nœud racine jusqu'au nœud partitionné. Pour limiter l'exploration des nœuds redondants, nous avons proposé d'ajouter un seuil de partitionnement qui limite le nombre des sous-arbres partitionnés. Ce seuil est augmenté automatiquement à chaque fois qu'un déséquilibre de charge entre les cœurs de calcul est détecté~\cite{TMBLPCO13}.

\begin{figure}[htbp]
\begin{minipage}[c]{.35\linewidth}
\includegraphics[scale=0.19]{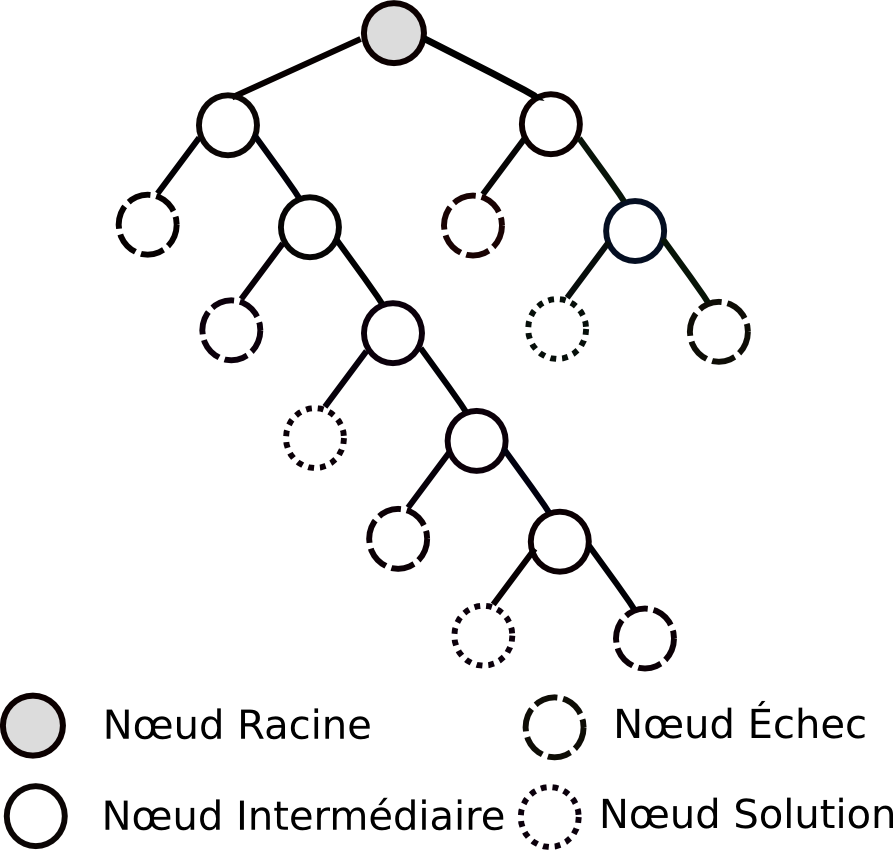}
\caption {Arbre de Recherche généré par le solveur OR-Tools} \label{ort_arbre}
\end{minipage}
\hfill
\begin{minipage}[c]{.5\linewidth}
\includegraphics[scale=0.165]{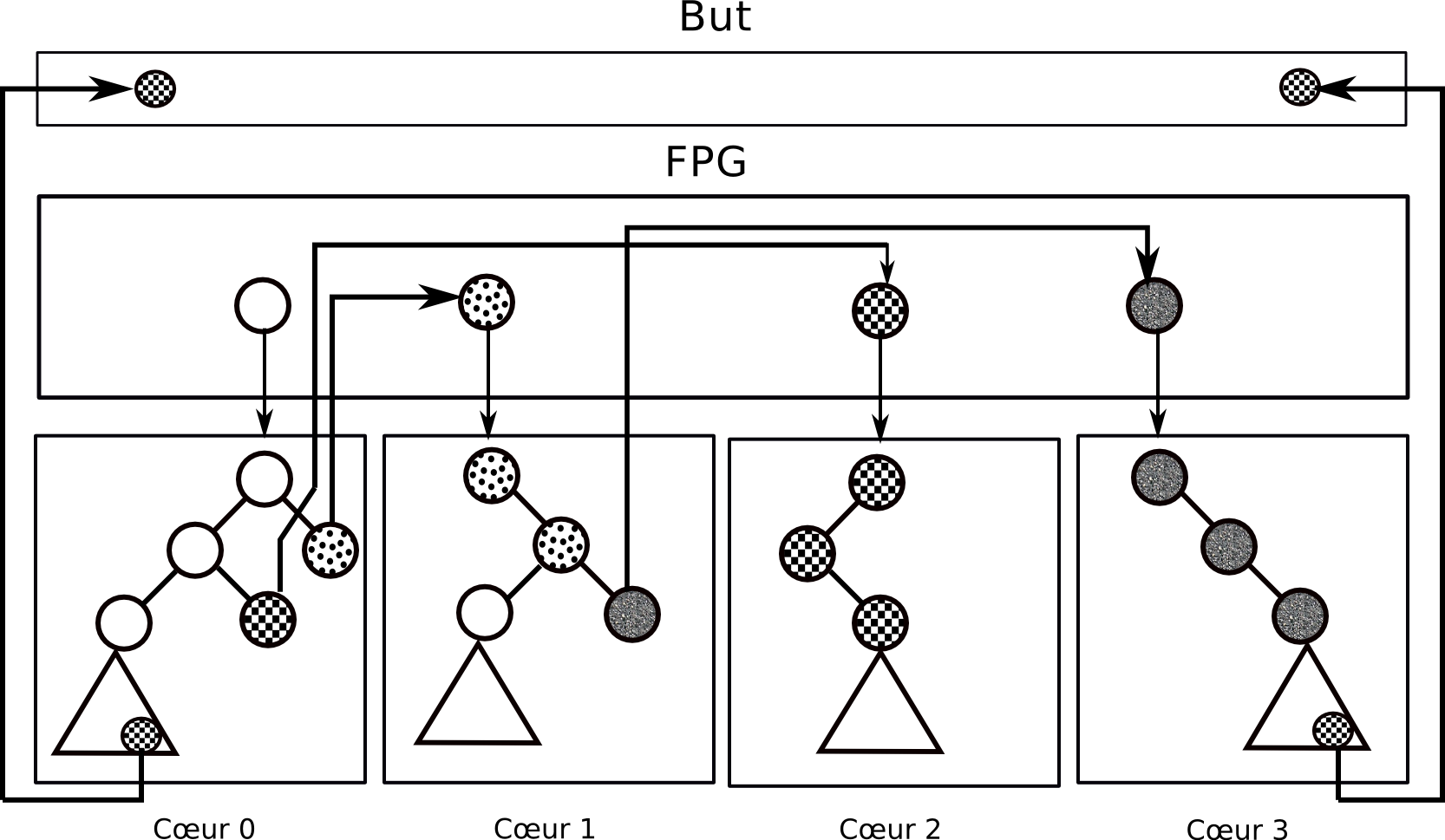}
\caption {Recherche Parallèle effectuée par la Stratégie de Partitionnement Déterministe} \label{parallel_SPD}
\end{minipage}
\end{figure}

L'objectif de la SPDA est d'explorer tout l'arbre de recherche pour trouver toutes les solutions possibles. C'est pour cela que le \textit{nœud-BOB} représente le chemin du nœud racine vers le nœud droit le plus élevé afin de minimiser le nombre de nœuds redondants.
Au contraire, la SPD n'est pas obligée d'explorer tous les nœuds de l'arbre de recherche, car le but est de trouver la première solution ou la première solution optimale la plus à gauche.
Dans le cadre des arbres de recherche générés par le solveur OR-Tools, les nœuds solutions sont les nœuds feuilles tels que présentés dans la figure~\ref{ort_arbre}, ce qui signifie que les nœuds solutions sont les nœuds les plus bas de l'arbre de recherche.
Par conséquent, en utilisant la SPD, \textbf{le \textit {nœud-BOB} représente le chemin du nœud racine jusqu'au nœud droit le plus bas}, afin d'explorer d'abord les nœuds les plus proches des nœuds solution.

La figure~\ref{parallel_SPD} montre la progression d'une exploration parallèle d'un espace de recherche de PPC en utilisant la SPD.

\begin{itemize}
\item Le cœur 0 commence la recherche avec le nœud racine, il détecte qu'il existe des cœurs en attente (cœur 1, 2 et 3). Dans ce cas, le cœur 0 insère le nœud droit le plus bas dans la FPG,
\item Le cœur 1 prend le nœud inséré dans la FPG et commence la recherche. Chacun des deux cœurs actifs (cœur 0 et 1) détecte en parallèle l'existence des cœurs en attente (cœur 2 et 3). Donc, chaque cœur actif insère dans la FPG le nœud droit le plus bas de son sous-arbre.
\item Tous les cœurs de calcul travaillent en parallèle, ce qui facilite l'équilibrage de charge entre les cœurs.
\item Quand une solution est trouvée, elle est sauvegardée dans un espace global appelé dans le framework Bobpp le \textit{But}, qui est un espace partagé entre tous les cœurs de calcul.
\end{itemize}

Pour retourner toujours la même solution en utilisant le mode parallèle, la SPD utilise un identifiant unique pour chaque nœud solution. Comme présenté dans la section~\ref{intro}, l'identifiant de nœud représente le chemin du nœud racine au nœud solution. Pour trouver la même solution avec n'importe quel nombre de cœurs, la SPD change le comportement en fonction du type de problème de PPC. Les sous-sections~\ref{SPD_csp} et~\ref{SPD_cop} présentent le comportement de la SPD pour résoudre les PSC et les POC.

\subsection{Stratégie de Partitionnement Déterministe pour Résoudre les Problèmes de Satisfaction de Contraintes}\label{SPD_csp}

Pour résoudre un PSC avec un algorithme déterministe, nous ajoutons quelques contraintes pour les cœurs actifs utilisés dans la recherche parallèle. Donc, à chaque fois qu'un cœur trouve une solution, il la sauvegarde comme étant la Solution la Plus à Gauche (SPG) et demande aux autres cœurs qui explorent en parallèle les branches qui se situent à droite de la branche de la SPG d'arrêter leurs recherches afin de concentrer la recherche sur la partie gauche de la branche de la SPG. La SPG est partagée entre tous les cœurs de calcul. Si une nouvelle solution est trouvée, la SPG est mise à jour par cette nouvelle solution. Nous effectuons cette mise à jour parce que l'identifiant de la nouvelle solution est toujours située à gauche de l'identifiant de la SPG. Lorsque l'exploration parallèle de l'arbre de recherche est terminée, la SPD retourne à l'utilisateur la SPG qui est la même solution que la première solution obtenue en séquentiel.

\begin{algorithm}[t]
\begin{algorithmic}
\REQUIRE $\mathcal{S}$, le seuil de partitionnement
\STATE $P$, le seuil de nœud courant 
\STATE $SPG$, la Solution la Plus à Gauche, initialement égale à nulle
\IF{une solution $S$ est trouvée}
        \IF{$S$ est plus à gauche que $SPG$ }
        		\STATE $SPG$=$S$
	        \STATE Arrêter tous les cœurs qui travaillent à droite de la branche de la $SPG$
        \ENDIF
\ELSE
    \IF {$\exists$ au moins un cœur en attente et $P<S$}
        \STATE Arrêter la recherche sur le nœud gauche
        \STATE Créer le \textit{nœud-BOB} (sous-arbre droit) et l'insérer dans la FPG
        \STATE Continuer la recherche avec le nœud gauche
	\ELSE
     	\STATE Effectuer une exploration séquentielle de l'espace de recherche
    \ENDIF
\ENDIF  
\end{algorithmic}
\caption{Algorithme de Recherche Parallèle Déterministe}
\label{Algo1}
\end{algorithm}

\begin{figure}[htbp]
\begin{minipage}[c]{.5\linewidth}
\center \includegraphics[scale=0.37]{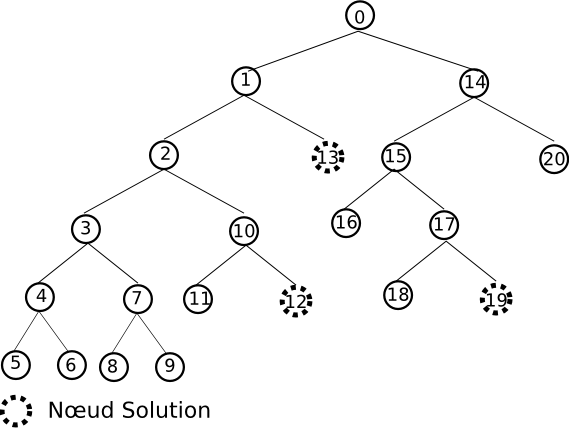}
\caption{Arbre de Recherche pour Résoudre un Problème de Satisfaction de Contraintes}
\label{fig_arbre}

\end{minipage}
\hfill
\begin{minipage}[c]{.45\linewidth}
\small
        \center \begin{tabular}{|c|c|c|c|}
        \hline
        Étapes & Cœur 0 & Cœur 1 & Cœur 2\\ \hline
        1 & Nœud 1 & \textit{En Attente} & \textit{En Attente} \\ \hline
        2 & Nœud 2 & Nœud 14 & \textit{En Attente} \\ \hline
        3 & Nœud 3 & Nœud 15 & Nœud 13 \\ \hline
        4 & Nœud 4 & \textit{En Attente} & \textit{En Attente} \\ \hline
        5 & Nœud 5 & Nœud 7 & Nœud 10 \\ \hline 
        6 & Nœud 6 & Nœud 8 & Nœud 11 \\ \hline                                                 
        7 & \textit{En Attente} & Nœud 9 & Nœud 12\\ \hline
        \end{tabular}
        \caption {Un Ordonnancement des cœurs de calcul utilisé pour résoudre un Problème de Satisfaction de Contraintes}
         \label{table_1}

\end{minipage}
\end{figure}

L'algorithme~\ref{Algo1} montre le principe de la SPD pour résoudre un PSC qui est représenté par son arbre de recherche dans la figure~\ref{fig_arbre}.

Pour trouver toujours la même solution représentée par le nœud 12 dans la figure~\ref{fig_arbre} en utilisant la SPD avec 3 cœurs, nous exécutons les étapes suivantes. Pour aider le lecteur à suivre les progrès de la SPD, nous n'avons pas pris en compte le parcours redondant des nœuds qui est expliqué en détail dans~\cite{TMBLPCO13} :

\begin{itemize}
\item Étape 0 : Le cœur 0 commence la recherche avec le nœud racine, 
\item Étape 1 : Le cœur 0 visite le nœud 1 et il détecte qu'il existe d'autre cœurs en attente (cœur 1 et 2), donc le cœur 0 insère le nœud 14 dans la FPG,
\item Étape 2 : Le cœur 0 visite le nœud 2 et insère le nœud 13 dans la FPG car le cœur 2 est toujours en attente, le cœur 1 visite le nœud 14,
\item Étape 3 : Le cœur 0 visite le nœud 3, le cœur 1 visite le nœud 15 et le cœur 2 visite le nœud 13 qui est un nœud solution. Comme le cœur 1 visite le nœud 15 qui est à droite de nœud solution, le cœur 2 demande au cœur 1 d'arrêter la recherche,
\item Étape 4 : Le cœur 0 visite le nœud 4 et détecte que les cœurs 1 et 2 sont en attente. Le cœur 0 insère les nœuds 7 et 10 dans la FPG. Le cœur 1 est en attente car il a été arrêté par le cœur 2. Le cœur 2 est en attente parce qu'il a terminé sa recherche dans son sous-arbre.
\item Étape 5 : Le cœur 0 visite le nœud 5, le cœur 1 visite le nœud 7 et le cœur 2 visite le nœud 10,
\item Étape 6 : Le cœur 0 visite le nœud 6, le cœur 1 visite le nœud 8 et le cœur 2 visite le nœud 11,
\item Étape 7 : Pour finir, le cœur 0 est en attente, le cœur 1 visite le nœud 9 et le cœur 2 visite le nœud 12. Le nœud 12 est un nœud solution et comme il n'y a pas de cœur qui explore la recherche à droite de ce nœud solution, le cœur 2 ne demande pas aux autres cœurs d'arrêter la recherche.
\end{itemize}

Le tableau~\ref{table_1} est un résumé d'un ordonnancement des cœurs de calcul utilisé par la SPD pour résoudre un PSC. Lorsque la SPD termine l'exploration de l'arbre de recherche, il retourne à l'utilisateur la solution la plus à gauche, représentée dans la figure~\ref{fig_arbre} par le nœud 12. Cette solution est la même que la première solution retournée en utilisant l'exécution séquentielle.

\subsection{Stratégie de Partitionnement Déterministe pour Résoudre les Problèmes d'Optimisation de Contraintes}\label{SPD_cop}

La différence entre les PSC et les POC est que dans les PSC toutes les solutions ont la même évaluation, alors que dans les POC, les solutions n'ont pas toutes la même évaluation. Ainsi résoudre un PSC consiste simplement à trouver une solution, et résoudre un POC consiste à trouver une solution dont la valeur est optimale.

\begin{figure}[htbp]
\begin{minipage}[c]{.5\linewidth}
\center \includegraphics[scale=0.44]{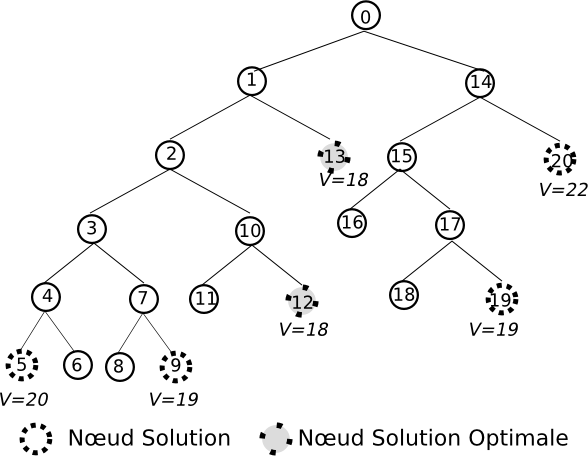}
\caption{Arbre de Recherche pour Résoudre un Problème d'Optimisation de Contraintes}
\label{fig_arbre3}
\end{minipage}
\hfill
\begin{minipage}[c]{.45\linewidth}
\small
 \center \begin{tabular}{|c|c|c|c|}
        \hline
        Étapes & Cœur 0 & Cœur 1 & Cœur 2\\ \hline
        1 & Nœud 1 & \textit{En Attente} & \textit{En Attente} \\ \hline
        2 & Nœud 2 & Nœud 14 & \textit{En Attente} \\ \hline
        3 & Nœud 3 & Nœud 15 & Nœud 13 \\ \hline
        4 & Nœud 4 & Nœud 16 & \textit{En Attente} \\ \hline
        5 & Nœud 5 & Nœud 17 & Nœud 7 \\ \hline 
        6 & Nœud 6 & Nœud 18 & Nœud 8 \\ \hline                                                 
        7 & Nœud 10 & Nœud 19 & Nœud 9\\ \hline
        8 & Nœud 11 & Nœud 20 & \textit{En Attente}\\ \hline
        9 & Nœud 12 & \textit{En Attente} & \textit{En Attente}\\ \hline
        \end{tabular}
        \caption{Un Ordonnancement des cœurs de calcul utilisé pour résoudre un Problème d'Optimisation de Contraintes}        
         \label{table_2}     

\end{minipage}
\end{figure}

Par exemple, dans la figure~\ref{fig_arbre3}, qui représente l'arbre de recherche généré pour résoudre un POC (cas d'un problème de minimisation), nous pouvons voir qu'il y a beaucoup de solutions, mais la solution la plus optimale (avec une évaluation minimum) est la solution représentée par le nœud 12 et qui a la plus petite évaluation (18). Dans un espace de recherche généré pour résoudre un POC, il est possible de trouver plusieurs solutions optimales. Dans l'exécution séquentielle, à chaque fois que l'algorithme de recherche trouve une solution, il évalue et sauvegarde la solution la plus optimale. Lorsque l'algorithme de recherche visite tous les nœuds, il retourne à l'utilisateur la première solution la plus optimale. Pour obtenir cette même solution optimale en utilisant une exécution parallèle, nous utilisons le même principe que la SPD pour résoudre les PSC avec une petite modification. Quand un cœur actif trouve une solution, il la sauvegarde en tant que la Solution Optimale la Plus à Gauche (SOPG) et il n'oblige pas les autres cœurs actifs à arrêter leur recherche. Si une nouvelle solution est trouvée, la SOPG est mise à jour dans deux cas :

\begin{itemize}
\item Premier cas : L'évaluation de la nouvelle solution est plus optimale que l'évaluation de la SOPG,
\item Second cas : La SOPG et la nouvelle solution ont la même évaluation, mais l'identifiant de la nouvelle solution est plus à gauche que l'identifiant de la SOPG.
\end{itemize}

Pour trouver toujours la même solution représentée par le nœud 12 dans la figure~\ref{fig_arbre3}, en utilisant la SPD avec 3 cœurs, nous exécutons les étapes suivantes :

\begin{itemize}
\item Étapes 0, 1 et 2 : Mêmes étapes que celles vues pour résoudre un PSC,
\item Étape 3 : Le cœur 0 visite le nœud 3, le cœur 1 visite le nœud 15 et le cœur 2 visite le nœud 13. Le nœud 13 est un nœud solution, il est sauvegardé comme étant la solution optimale la plus à gauche,
\item Étape 4 : Le cœur 0 visite le nœud 4 et détecte que le cœur 2 est en attente. Le cœur 0 insère le nœud 7 dans la FPG. Le cœur 1 visite le nœud 16,
\item Étape 5 : Le cœur 0 visite le nœud 5 qui est un nœud solution, mais nous ne pouvons pas mettre à jour la solution optimale la plus à gauche avec cette solution car l'évaluation du nœud 5 est égale à 20 ce qui est plus grand que l'évaluation de la solution optimale la plus à gauche (valeur 18 ). Le cœur 1 visite le nœud 17 et le cœur 2 visite le nœud 7,
\item Étape 6 : Le cœur 0 visite le nœud 6, le cœur 1 visite le nœud 18 et le cœur 2 visite le nœud 8,
\item Étape 7 : Le cœur 0 visite le nœud 10, le cœur 1 visite le nœud 19 et le cœur 2 visite le nœud 9. Les nœuds 9 et 19 sont des nœuds solutions avec la même évaluation (valeur 19), mais cette évaluation est plus grande que l'évaluation de la solution optimale la plus à gauche (valeur 18 ). Donc, nous ne pouvons pas mettre à jours la solution optimale la plus à gauche.
\item Étape 8 : Le cœur 0 visite le nœud 11, le cœur 1 visite le nœud 20 qui est un nœud solution avec une évaluation de 22 et le cœur 2 est en attente,
\item Pour finir, le cœur 0 visite le nœud 12, l'évaluation du nœud 12 est la même que l'évaluation de la solution optimale la plus à gauche, mais le nœud 12 est plus à gauche que le nœud de la solution optimale la plus à gauche. Dans ce cas, la solution optimale la plus à gauche est mise à jour par le nœud 12. Les cœurs 1 et 2 sont en attente.
\end{itemize}

Le tableau~\ref{table_2} est un résumé d'un ordonnancement des cœurs de calcul utilisé par la SPD pour résoudre un POC. Lorsque la SPD visite tous les nœuds, elle retourne à l'utilisateur la solution optimale la plus à gauche qui est représentée dans la figure~\ref{fig_arbre3} par le nœud 12, qui a une évaluation de 18. Cette solution optimale la plus à gauche est la même solution que celle de la première solution retournée en utilisant le mode séquentiel.

\section{Expérimentation}\label{xp}
Pour valider l'approche utilisée dans cette étude, les expériences ont été effectuées en utilisant une machine parallèle Intel Xeon X5650 (2.67 GHz) avec la technologie \textit{Hyper-Threading} (6 cœurs physiques pour 12 \textit{threads} de calcul) équipée de 48 Go de RAM, avec le système d'exploitation Linux. La version du solveur OR-Tools utilisée comme un moteur de recherche pour résoudre les problèmes de PPC est la version 2727. Le solveur OR-Tools génère un arbre de recherche binaire, mais la SPD proposée dans cet article peut être adaptée pour les arbres de recherche $ N$-aire. Tous les problèmes de PPC résolus sont modélisés en utilisant le format FlatZinc. Les temps de calcul présentés dans cette section sont donnés en secondes et sont une moyenne de plusieurs exécutions.

\begin{figure}[htbp]
\begin{minipage}[c]{.45\linewidth}
\includegraphics[scale=0.5]{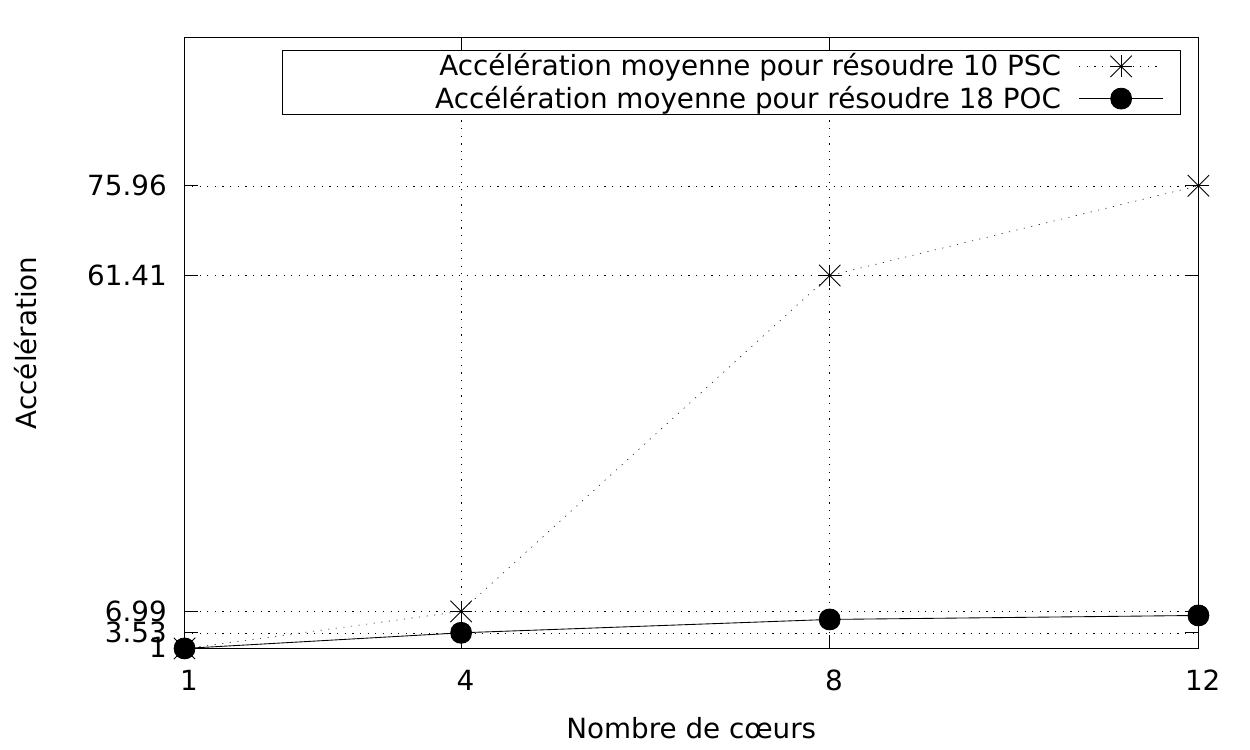}
\caption {Accélération moyenne pour trouver la première solution sans utiliser la fonctionnalité de déterminisme} \label{moy_sans_det}
\end{minipage}
\hfill
\begin{minipage}[c]{.45\linewidth}
\includegraphics[scale=0.55]{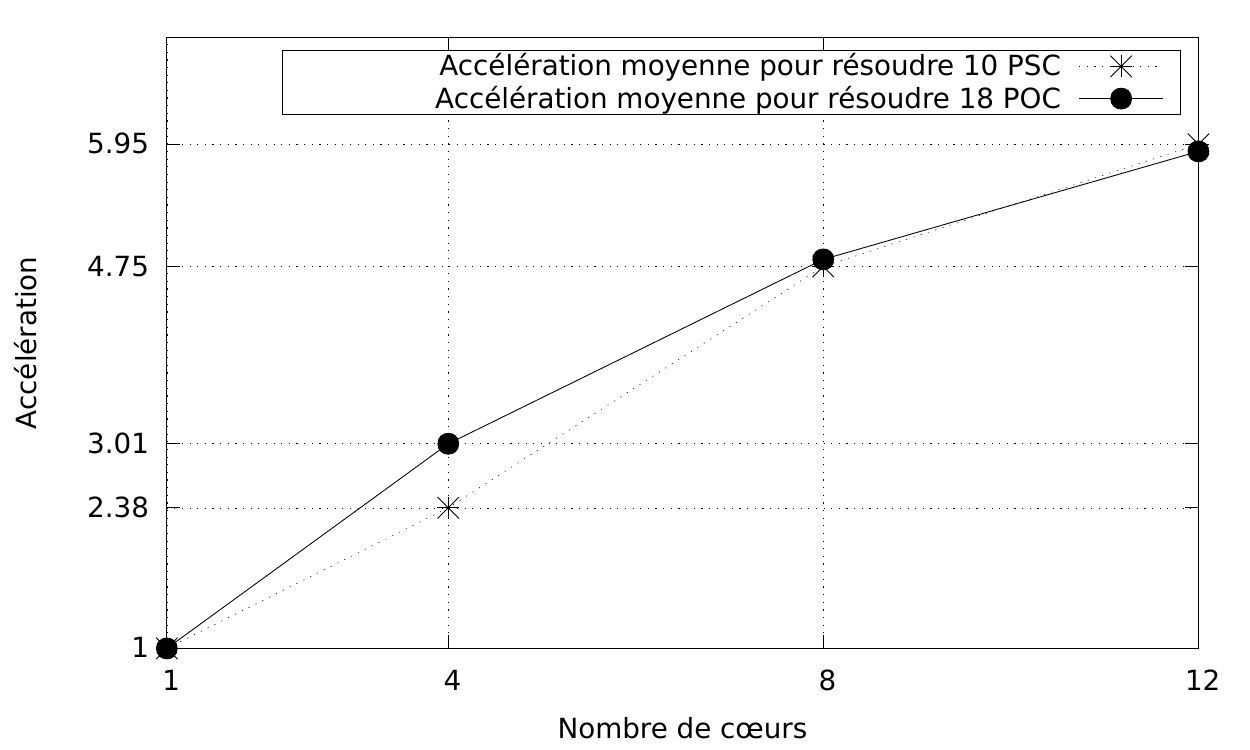}
\caption {Accélération moyenne pour trouver la première solution en utilisant la fonctionnalité de déterminisme} \label{moy_avec_det}
\end{minipage}
\end{figure}

La figure~\ref{moy_sans_det} \textit{(resp. figure~\ref{moy_avec_det})} représente l'accélération moyenne pour résoudre un ensemble de problèmes de PPC sans utiliser la fonctionnalité de déterminisme \textit{(resp. avec la fonctionnalité de déterminisme)}. Dans la figure~\ref{moy_sans_det}, pour résoudre les PSC nous avons cherché à trouver la première solution réalisable, par contre pour résoudre les POC nous avons cherché à trouver la première solution la plus optimale. Les problèmes résolus sont issus du MiniZinc Challenge 2012~\cite{ch2012}. Le but de ce défi est de comparer les solveurs de PPC et les différentes méthodes de PPC utilisées pour résoudre le même problème. Nous avons résolu dans ces deux expériences 28 problèmes de PPC, 10 sont des PSC et 18 sont des POC.

En conséquence, pour résoudre les problèmes de PPC, la fonctionnalité de déterminisme permet d'assurer la même solution quel que soit le mode d'exécution séquentiel ou parallèle, au prix de perdre en performance par rapport aux performances obtenues sans utiliser le déterminisme.

L'exécutions de la SPD sur une machine parallèle avec 12 cœurs de calcul, nous a permis d'atteindre une accélération de 5.95 pour résoudre les PSC et une accélération de 5.88 pour résoudre les POC.

\begin{figure}[htbp]
\begin{minipage}[c]{.45\linewidth}
\begin{center}
 \includegraphics[scale=0.34]{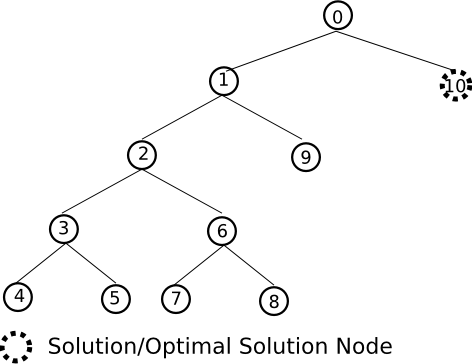}
\caption {Arbre de recherche donnant une bonne accélération} \label{fig_arbre5}
\end{center}
\end{minipage}
\hfill
\begin{minipage}[c]{.45\linewidth}
\begin{center}
 \includegraphics[scale=0.34]{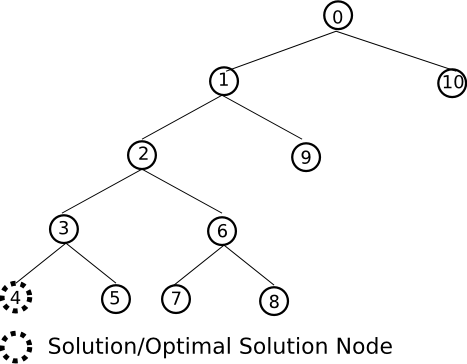}
\caption {Arbre de recherche donnant une mauvaise accélération} \label{fig_arbre4}
\end{center}
\end{minipage}
\end{figure}

\begin{figure}[htbp]
\begin{minipage}[c]{.45\linewidth}
\includegraphics[scale=0.55]{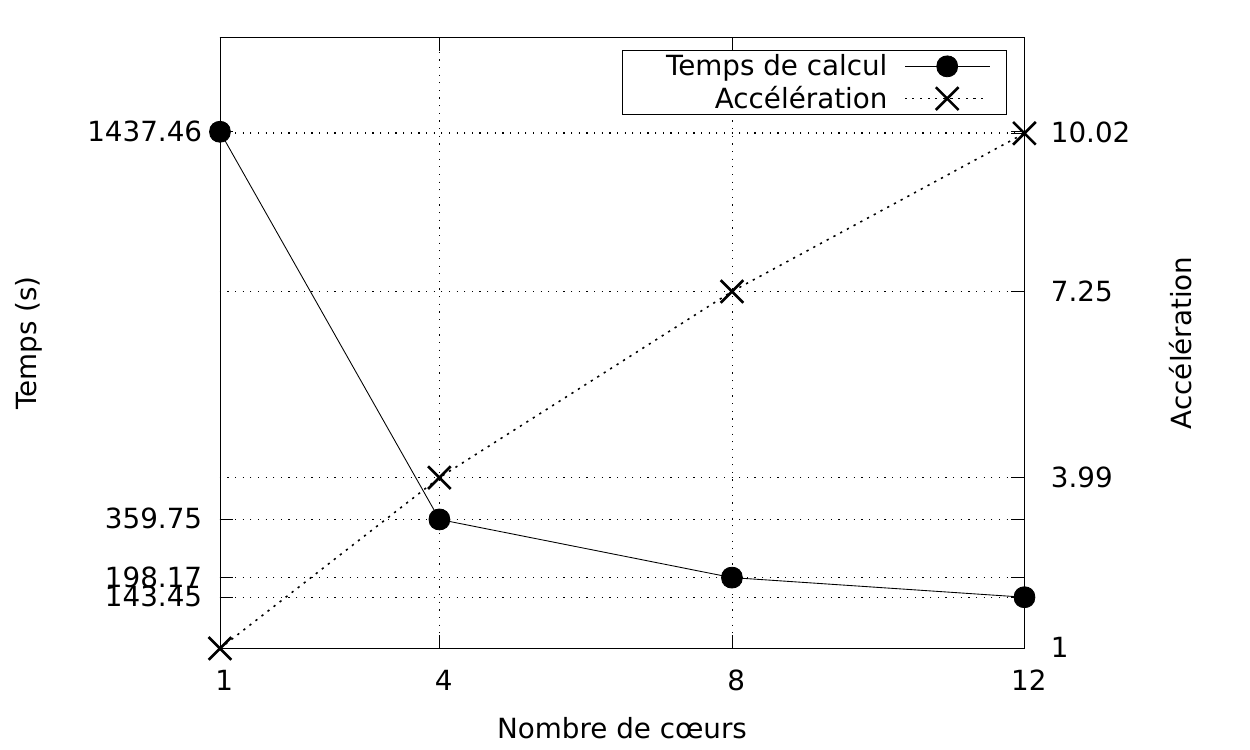}
\caption {Temps de calcul et accélération pour résoudre le problème de Quasi Groupe (Quasigroup7\_10~\cite{ch2012}) qui est un PSC} \label{max_sta}
\end{minipage}
\hfill
\begin{minipage}[c]{.45\linewidth}
\includegraphics[scale=0.55]{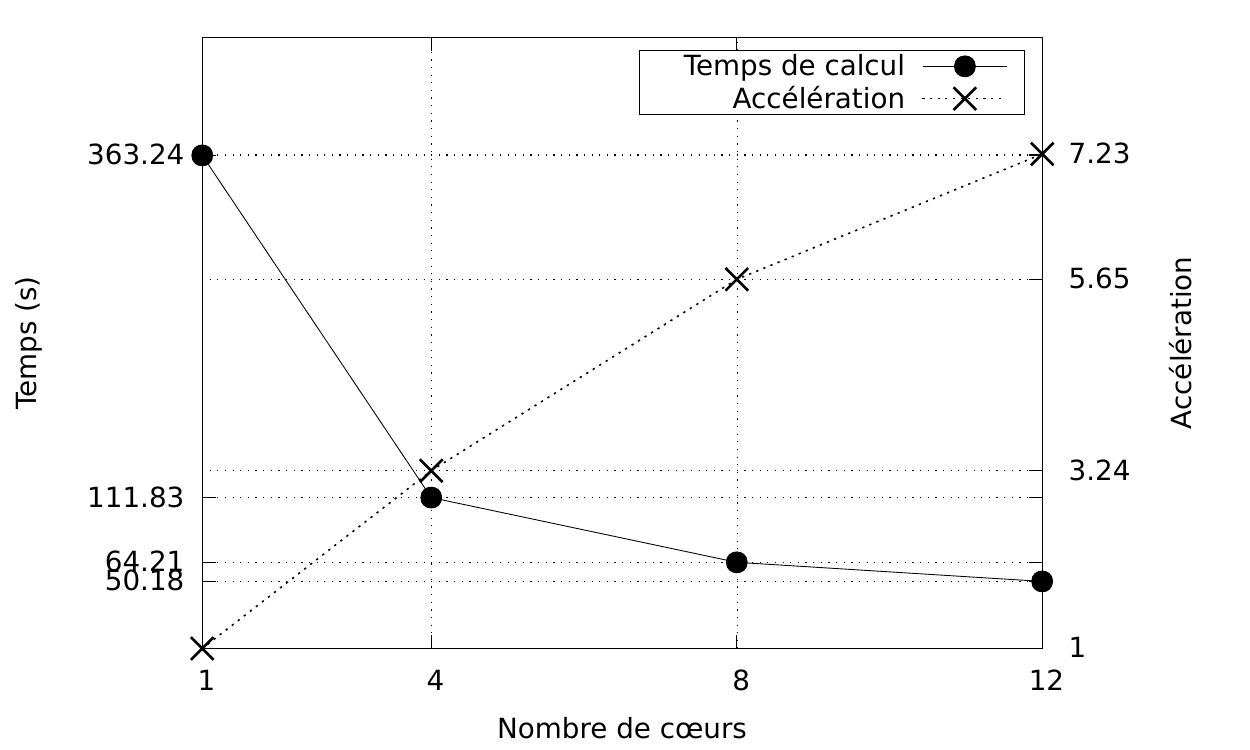}
\caption {Temps de calcul et accélération pour résoudre le problème Fast Food  (fastfood\_ff63~\cite{ch2012}) qui est un POC} \label{max_optim}
\end{minipage}
\end{figure}

Comme la SPD n'explore pas tout l'espace de recherche, la première solution ou la première solution optimale peut être située dans la branche droite de premier niveau de l'arbre de recherche, comme présenté dans la figure~\ref{fig_arbre5}. Dans ce cas, la SPD donne une bonne accélération \textit{(Best case)}. Les figures~\ref{max_sta} et~\ref{max_optim} confirment cet effet. 

\begin{figure}[htbp]
\begin{minipage}[c]{.45\linewidth}
\includegraphics[scale=0.55]{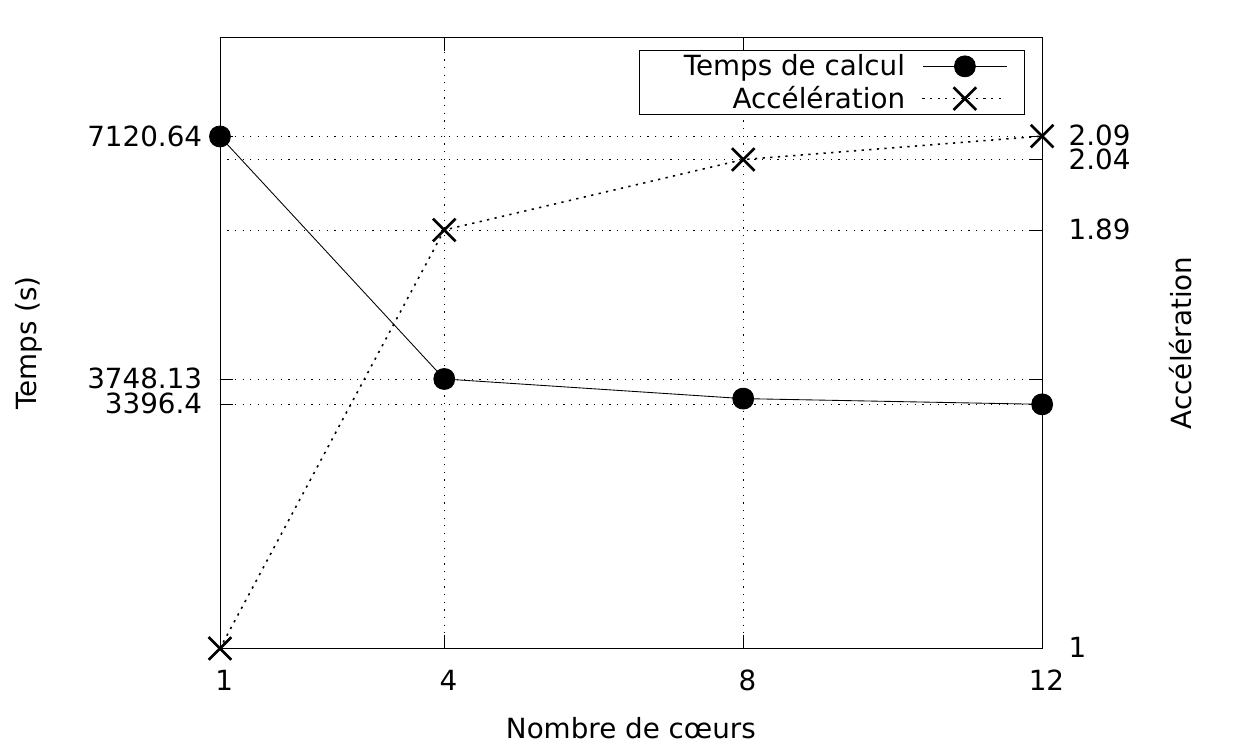}
\caption {Temps de calcul et accélération pour résoudre le problème Market (market\_split\_s5\-03~\cite{ch2012}) qui est un PSC} \label{min_sta}
\end{minipage}
\hfill
\begin{minipage}[c]{.45\linewidth}
\includegraphics[scale=0.55]{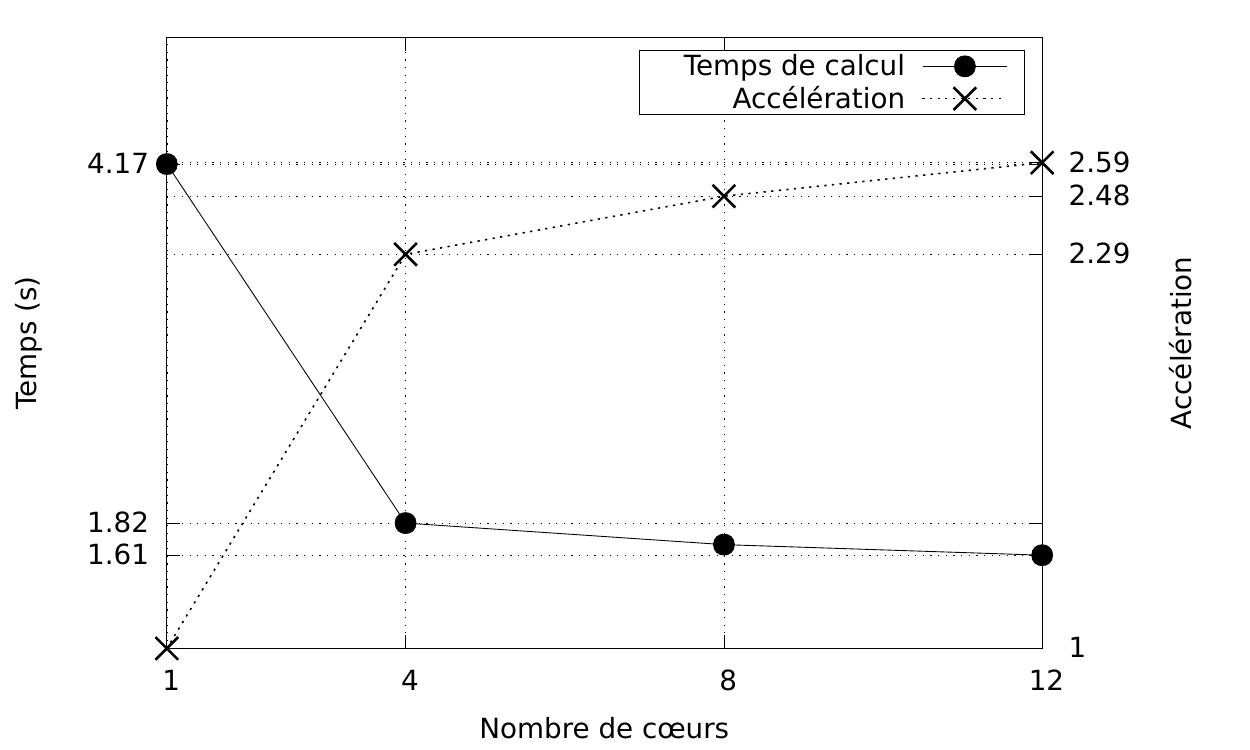}
\caption {Temps de calcul et accélération pour résoudre le problème de Fast Food (fastfood\_ff71~\cite{ch2012}) qui est un POC} \label{min_optim}
\end{minipage}
\end{figure}

Parfois, la première solution ou la première solution optimale est représentée par le nœud gauche le plus bas, tel que présenté dans la figure~\ref{fig_arbre4}. Dans ce cas, la SPD ne donne pas une bonne accélération \textit{(Worst case)}. Les figures~\ref{min_sta} et~\ref{min_optim} confirment cet effet.


\begin{figure}[htbp]
\begin{minipage}[c]{.45\linewidth}
\includegraphics[scale=0.53]{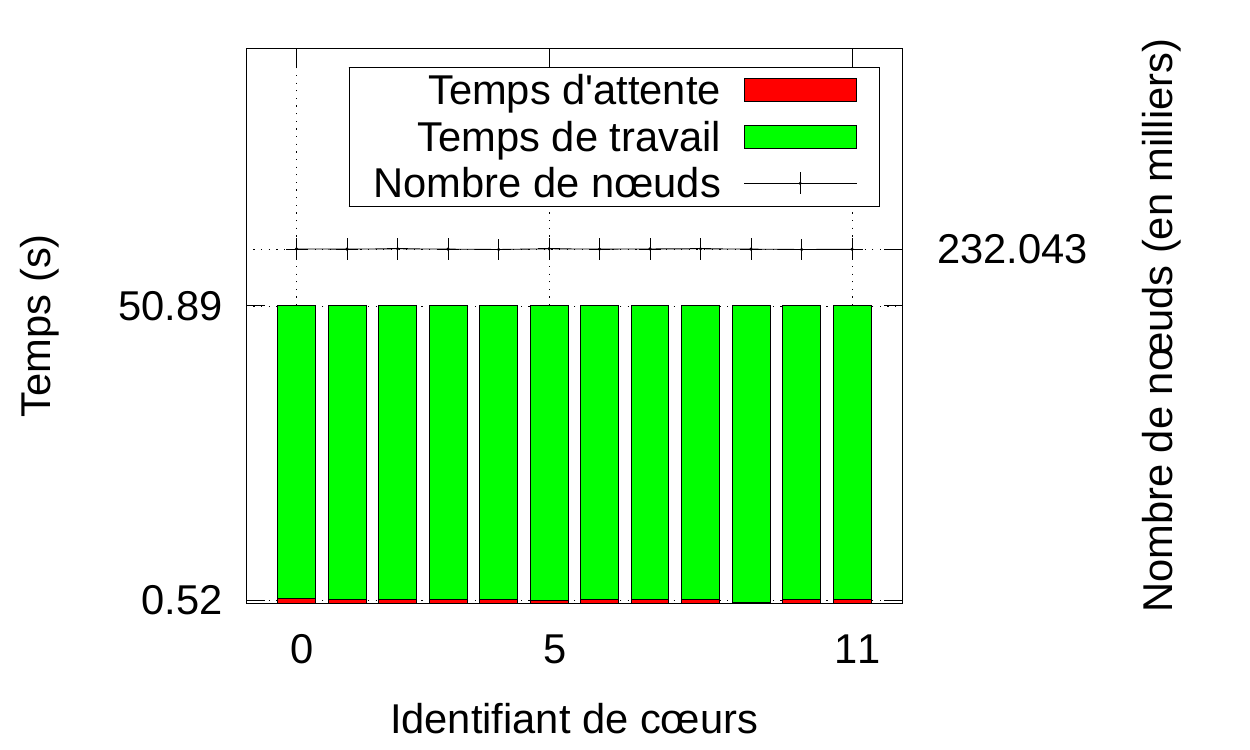}
\caption {Équilibrage de charge pour résoudre le problème de Naval Battle (Sb\_sb\_13\_13\_5\_1~\cite{ch2012}) qui est un PSC en utilisant 12 cœurs de calcul} \label{lb_sta}
\end{minipage}
\hfill
\begin{minipage}[c]{.45\linewidth}
\includegraphics[scale=0.53]{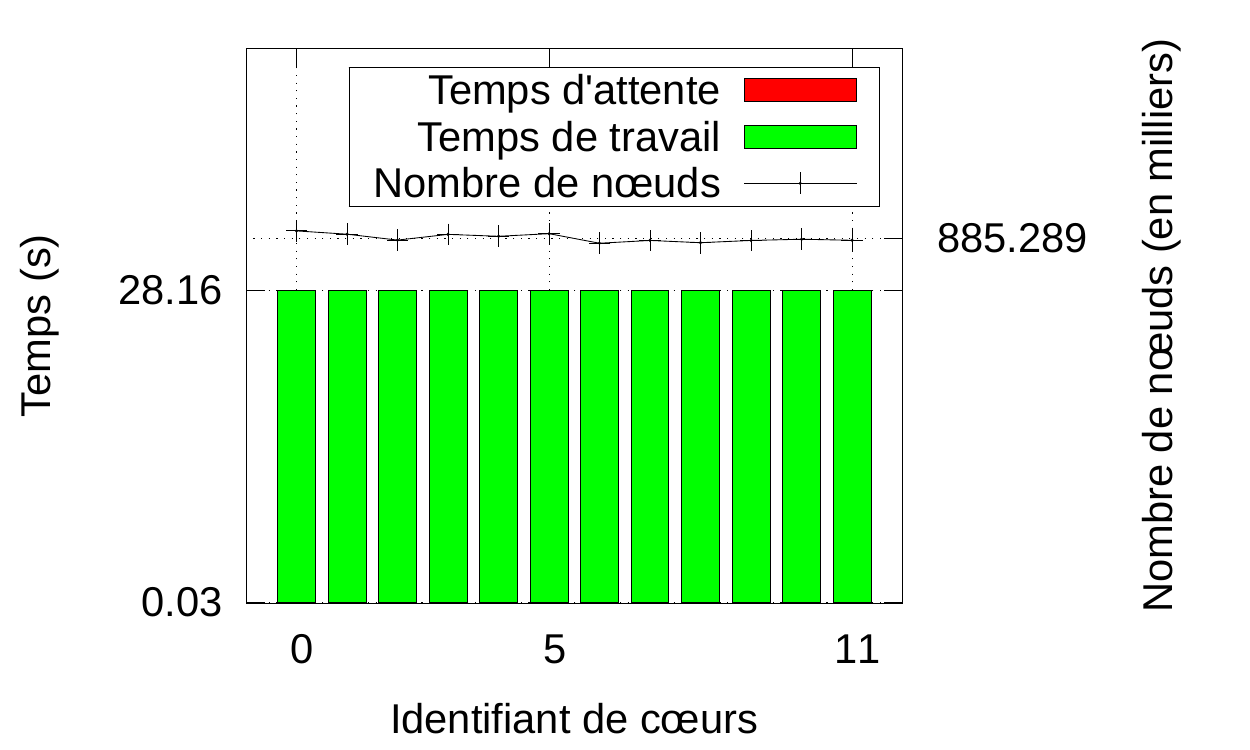}
\caption {Équilibrage de charge pour résoudre le problème de Fast Food (fastfood\_ff61~\cite{ch2012}) qui est un POC en utilisant 12 cœurs de calcul} \label{lb_optim}
\end{minipage}
\end{figure}

Les figures~\ref{lb_sta} et~\ref{lb_optim} montrent le comportement des cœurs de calcul lorsque nous utilisons la SPD. Comme résultat, les temps de calcul et d'attente sont répartis de manière équitables entre les cœurs de calcul. De plus, tous les cœurs ont visité le même nombre de nœuds OR-Tools.

\section{Conclusion et Perspectives}\label{conclusion}
Cet article présente une nouvelle stratégie de partitionnement déterministe parallèle qui retourne toujours à l'utilisateur la même solution en utilisant le mode séquentiel ou parallèle.

L'utilisation du framework parallèle Bobpp donne une bonne accélération sur des architectures à mémoire partagée. Ces résultats sont obtenus avec plusieurs types de problèmes d'optimisation combinatoire sur différents ordinateurs, Bobpp a également une parallélisation sur des architectures à mémoire distribuée version mixte MPI/Pthreads. Une première perspective est d'adapter cette Stratégie de Partitionnement Déterministe pour retourner à l'utilisateur toujours la même solution sur une architecture à mémoire distribuée.

Il est également intéressant d'enrichir et d'étendre l'ensemble des problèmes résolus en utilisant le framework Bobpp. Comme deuxième perspective, on peut porter au-dessus du framework Bobpp un solveur de SATisfiabilité booléenne (SAT), comme le solveur \textit{Glucose}~\cite{glucose}.

\bibliography{biblio}

\def\No{\kern-.25em\lower.2ex\hbox{\char'27}}
\begin{thebibliography}{10}

\bibitem{11HYSM09}
Alejandro (A.), Youssef (H.) et Mich{\`e}le (S.). --
\newblock {Online Heuristic Selection in Constraint Programming}, 2009.
  International Symposium on Combinatorial Search - 2009.

\bibitem{BJRE75}
Bitner, R. (J.), Reingold et M. (E.). --
\newblock Backtrack programming techniques. {\em Commun. ACM}, vol.~18, n\No{}
  11, novembre 1975, pp. 651--656.

\bibitem{BAAS09}
Bocchino, Jr. (R.~L.), Adve (V.~S.), Adve (S.~V.) et Snir (M.). --
\newblock Parallel programming must be deterministic by default. --
\newblock In {\em Proceedings of the First USENIX conference on Hot topics in
  parallelism}, {\em HotPar'09}, HotPar'09, pp. 4--4, Berkeley, CA, USA, 2009.
  USENIX Association.

\bibitem{DJLBCLOM09}
Devietti (J.), Lucia (B.), Ceze (L.) et Oskin (M.). --
\newblock Dmp: deterministic shared memory multiprocessing. {\em SIGARCH
  Comput. Archit. News}, vol.~37, n\No{} 1, mars 2009, pp. 85--96.

\bibitem{GCCSPS2009}
G.~Chu~\ (C.~S.) et Stuckey (P.). --
\newblock Confidence-based work stealing in parallel constraint programming.
  {\em In Principles and Practices of Constraint Programming}, 2009.

\bibitem{GaLe07}
Galea (F.) et {Le Cun} (B.). --
\newblock Bob++ : a framework for exact combinatorial optimization methods on
  parallel machines. --
\newblock In {\em International Conference High Performance Computing \&
  Simulation 2007 (HPCS'07) and in conjunction with The 21st European
  Conference on Modeling and Simulation (ECMS 2007)}, pp. 779--785, juin 2007.

\bibitem{glucose}
Glucose sat solver\\ http://www.labri.fr/perso/lsimon/glucose. Accessed:
  05-03-2014.

\bibitem{ibmilog}
Ibm ilog\\ http://www-01.ibm.com/software/info/ilog. Accessed: 05-03-2014.

\bibitem{TMBLPCO13}
Menouer (T.) et Cun (B.~L.). --
\newblock Anticipated dynamic load balancing strategy to parallelize constraint
  programming search. --
\newblock In {\em 2013 IEEE 27th International Symposium on Parallel and
  Distributed Processing Workshops and PhD Forum}, pp. 1771--1777, 2013.

\bibitem{MLSAVP2007}
Michel (L.), See (A.) et Hentenryck (P.). --
\newblock Parallelizing constraint programs transparently. {\em In\,: }{\em
  Principles and Practice of Constraint Programming – CP 2007}, \'ed. par
  Bessière (C.), pp. 514--528. --
\newblock Springer Berlin Heidelberg, 2007.

\bibitem{LaSH09}
Michel (L.), See (A.) et Van~Hentenryck (P.). --
\newblock Transparent parallelization of constraint programming. {\em INFORMS
  JOURNAL ON COMPUTING}, vol.~21, n\No{} 3, 2009, pp. 363--382.

\bibitem{ch2012}
Minizinc challenge\\
  \verb|http://www.minizinc.org/challenge2012/challenge.html|. Accessed:
  05-03-2014.

\bibitem{MN}
Nielsen (M.). --
\newblock {\em Parallel Search in Gecode}. --
\newblock Rapport technique, Gecode, 2006.

\bibitem{OMAJAS09}
Olszewski (M.), Ansel (J.) et Amarasinghe (S.). --
\newblock Kendo: efficient deterministic multithreading in software. {\em
  SIGARCH Comput. Archit. News}, vol.~37, n\No{} 1, mars 2009, pp. 97--108.

\bibitem{Pajero}
Pajero project\\ http://www.horizontalsoftware.com/fr/actualites/voir/31/.
  Accessed: 05-03-2014.

\bibitem{LP}
Perron (L.). --
\newblock Search procedures and parallelism in constraint programming. {\em
  Principles and Practices of Constraint Programming}, 1999.

\bibitem{IPCJIMNCPNPPCU}
P.Gent (I.), Jefferson (C.), Miguel (I.), Moore (N.~C.), Nightingale (P.),
  Prosser (P.) et Unsworth (C.). --
\newblock A preliminary review of literature on parallel constraint solving.
  {\em Proceedings PMCS'11 Workshop on Parallel Methods for Constraint
  Solving}, 2011.

\bibitem{WM04}
Refalo (P.). --
\newblock Impact-based search strategies for constraint programming. {\em In\,:
  }{\em Principles and Practice of Constraint Programming - CP 2004}, \'ed. par
  Wallace (M.), pp. 557--571. --
\newblock Springer Berlin Heidelberg, 2004.

\bibitem{Rolf11}
Rolf (C.~C.). --
\newblock {\em Parallelism in Constraint Programming}. --
\newblock Th\`ese de PhD, Department of Computer Science, Lund University, Oct
  2011.

\end{thebibliography}
\end{document}